\documentclass[showpacs,aps,prl,twocolumn,groupaddress]{revtex4-1}

\usepackage{graphicx}  
\usepackage{subfigure}
\usepackage{dcolumn}   
\usepackage{bm}        
\usepackage{bbold}        
\usepackage{amssymb}   
\usepackage{comment}
\usepackage{hyperref}
\usepackage{color}
\usepackage{amsmath}
\usepackage{mathrsfs}
\usepackage{mathcomp}
\usepackage{textcomp}
\usepackage{dsfont}
\usepackage{esint}
\usepackage{braket}
\usepackage{textgreek}
\usepackage{lipsum}
\usepackage{marvosym} 

\begin{document}

\title{Non-Hermitian exceptional Landau quantization in electric circuits}

\author{Xiao-Xiao Zhang}
\affiliation{Department of Physics and Astronomy \& Stewart Blusson Quantum Matter Institute, University of British Columbia, Vancouver, BC, V6T 1Z4 Canada}
\author{Marcel Franz}
\affiliation{Department of Physics and Astronomy \& Stewart Blusson Quantum Matter Institute, University of British Columbia, Vancouver, BC, V6T 1Z4 Canada}


\newcommand\dd{\mathrm{d}}
\newcommand\ii{\mathrm{i}}
\newcommand\ee{\mathrm{e}}
\newcommand\zz{\mathtt{z}}
\makeatletter
\let\newtitle\@title
\let\newauthor\@author
\let\newaddress\@affiliation
\newcommand\getauthor{\newauthor}
\def\ExtendSymbol#1#2#3#4#5{\ext@arrow 0099{\arrowfill@#1#2#3}{#4}{#5}}
\newcommand\LongEqual[2][]{\ExtendSymbol{=}{=}{=}{#1}{#2}}
\newcommand\LongArrow[2][]{\ExtendSymbol{-}{-}{\rightarrow}{#1}{#2}}
\newcommand{\cev}[1]{\reflectbox{\ensuremath{\vec{\reflectbox{\ensuremath{#1}}}}}}
\newcommand{\red}[1]{\textcolor{red}{#1}} 
\newcommand{\blue}[1]{}
\newcommand{\green}[1]{\textcolor{green}{#1}} 
\newcommand{\mycomment}[1]{} 
\makeatother

\begin{abstract}
Alternating current RLC electric circuits form an accessible and highly tunable platform simulating Hermitian as well as non-Hermitian (nH) quantum systems. We propose here a circuit realization of nH Dirac and Weyl Hamiltonians subject to time-reversal invariant pseudo-magnetic field, enabling the exploration of novel nH physics. We identify the low-energy physics with a generic real energy spectrum from the nH Landau quantization of exceptional points and rings, which can avoid the nH skin effect and provides a physical example of a quasiparticle moving in the complex plane. Realistic detection schemes are designed to probe the flat energy bands, sublattice polarization, edge states protected by a nH energy-reflection symmetry, and a characteristic nodeless probability distribution.
\end{abstract}

\maketitle



\textit{Introduction} --
Hermiticity of Hamiltonians has long been a required ingredient in any self-consistent framework of quantum theory for both the stationary and time-dependent problems. There has been an increasing effort aimed at understanding the phenomena resultant from relaxing the Hermiticity condition both as a theoretical challenge and as a description of various physical systems \cite{Moiseyev2009\mycomment{,nSelfAdjointBook2015}}. Theoretical and experimental efforts were largely ignited by the recognition of the $\mathcal{PT}$-symmetry \cite{Bender1998,Bender2007}, its realization in optics \cite{Makris2008,Guo2009,Rueter2010}, and further 
generalizations \cite{Mostafazadeh2003,Mostafazadeh2010}. In physical systems non-Hermiticity can arise through incorporating loss or gain but also by viewing Hermitian systems from new angles, including vortex pinning in superconductors \cite{Hatano1996\mycomment{,Hatano1997,Hatano1998}}, topological surface state \cite{Gonzalez2017,Molina2018}, and quasiparticles with self-energy correction \cite{Kozii2017,Shen2018a,Zyuzin2018,Papaj2019}. 
Important developments have recently been focused on the classification of new phases \cite{Leykam2017,Ashida2017,Shen2018,Kawabata2019,Gong2018,Alvarez2018,Lin2019,Zhang2019,Yang2019,Wang2019,Ghatak2019,Torres2019}, the bulk Fermi arc and line structures \cite{Kozii2017,Papaj2019,Zhou2018,Carlstroem2018} and the anomalous bulk-boundary correspondence with the skin effect where macroscopically many states are localized at the boundary \cite{Lee2016,Alvarez2018a,Xiong2018,Kunst2018,Yao2018a,Yao2018,Yokomizo2019,Jin2019,Lee2019,Ezawa2019a,Helbig2019a,Xiao2019}. 
Of particular importance are the 
generic exceptional degeneracies -- exceptional point (EP) in two dimensions (2D) and exceptional ring (ER) in three dimensions (3D) -- in the complex energy spectrum where two resonances match at once in position and width \cite{Berry2004,Heiss2012,Xu2017,Cerjan2018,Okugawa2019,Luo2018}.  Signatures have been experimentally observed in microwave cavities \cite{Dembowski2001,Dembowski2004}, exciton-polariton systems \cite{Gao2015}, and photonic lattices \cite{Zhen2015,Cerjan2019}.

In this work we discuss a new family of phenomena arising from applying magnetic field to nontrivial non-Hermitian (nH) systems. This problem has remained largely unexplored owing to the lack of a feasible realization which we overcome here by considering a convenient synthetic platform based on alternating current (ac) circuits. Periodic arrays of capacitors and inductors are known to simulate the physics of electrons in crystal lattices and can model various topological phases \cite{Ningyuan2015,Albert2015,Zhu2018,Lee2018,Luo2018a,Zhao2018,Ezawa2018,Imhof2018,\mycomment{Luo2018,Ezawa2019,}Haenel2019,Ezawa2019b,\mycomment{Helbig2019,Hoffmann2019,}Yu2019,Lu2019,Serra-Garcia2019}. We introduce nH effects by including dissipative resistance in such arrays. 
Pseudo-magnetic fields (pMFs) can be generated by spatially varying certain electric elements, which extends to the nH case the pMF realized by elastic strain in relativistic electron systems \cite{Guinea2009,Levy2010,\mycomment{Vozmediano2010,}Cortijo2015,Pikulin2016,Peri2019,Rechtsman2012,Nigge2019}. 
The nH effects generically turn relativistic band crossings into exceptional degeneracies. Interplay with the pMF then results in a novel nH low-energy theory of \textit{bulk} states which have a \textit{real} energy spectrum and are free from the skin aggregation effect. In addition such systems exhibit novel edge states protected by strong nH energy-reflection symmetry and realize a physical analog of a particle moving in the complex domain. We explain how these remarkable phenomena can be detected via conventional electric measurements.

\textit{Circuit Realization} --
Based on the Kirchhoff current law (KCL), one can apply the node analysis to an ac circuit at frequency $\omega$. The Euler-Lagrange equation for the node flux variable $\varphi_j$ given the external current $i_j$ injected at node $j$ reads 
\begin{equation}\label{eq:EL_eq}
\frac{\dd}{\dd t}\frac{\partial L}{\partial \dot\varphi_j}-\frac{\partial L}{\partial \varphi_j}+\frac{\partial D}{\partial \dot\varphi_j}=i_j,
\end{equation}
where for capacitors and inductors $L_C=\frac{C}{2}\dot\varphi^2,L_L=-\frac{1}{2L}\varphi^2$ while Rayleigh dissipation function $D=\frac{1}{2R}\dot\varphi^2$ describes resistors. 
These equations form an admittance problem, $J \bm{v}=\bm{i}$, where the admittance matrix $J$ determines the voltage response $\bm v=\dot{\bm\varphi}$ in the circuit to an array of injected currents $\bm{i}=(i_1,i_2,\cdots,i_N)$.
At any fixed frequency $J$ can be mapped to a tight-binding Hamiltonian $H=-\ii J$ with hopping amplitudes $\omega C$ ($\frac{-1}{\omega L}$) for nodes connected by a capacitor (inductor) while a nH hopping $\ii /R$ accounts for any resistor and hence $H(R\mapsto-R)=H^\dag$.  Lossless LC circuits can fully simulate ordinary time-reversal ($\mathcal{T}$) invariant quantum Hamiltonians as the $\pi$-phase difference between $L,C$ hoppings suggests. This remains true in the presence of pMF which couples to Dirac/Weyl nodes in a way that respects $\mathcal{T}$. 
Solving the eigenproblem $H|\psi_\alpha\rangle= E_\alpha|\psi_\alpha\rangle$ corresponds to finding a spatial pattern of currents $\bm{i}_\alpha$ that produces the identical pattern of voltages $\ii E_\alpha\bm{v}_\alpha=\bm{i}_\alpha$. We discuss later how circuit tomography connects standard impedance or voltage measurements 
with the energy spectrum and the wavefunction of the quantum problem.
\begin{figure}[t]
\centering
\includegraphics[width=0.33\textwidth]{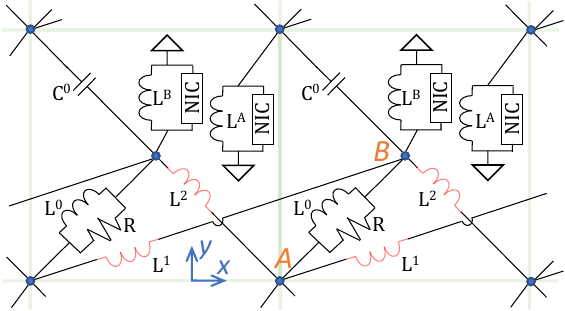}
\caption{RLC circuit 
simulating a quantum system with Dirac dispersions which turn into exceptional degeneracies upon inclusion of nH effects. pMF defined by gauge potential $A_y=bx$ can be generated by varying the red elements along the $x$-direction. Linear variation, required for uniform pMF, dictates open boundary condition along $x$.
}\label{Fig:circuit}
\end{figure}

Here we focus on  periodic circuits described by a family of two-band Bloch Hamiltonians $h(\bm{k}) = d_0\mathbb{1}+\sum_i{d_i\sigma_i}$ 
%
%
where psuedospin $\sigma$ stands for inequivalent nodes $A,B$. As described in Supplemental Material (SM) \cite{SM}, following the mapping between quantum model and ac KCL, 
a square-lattice circuit depicted in Fig.~\ref{Fig:circuit} 
can realize a variety of Hamiltonians of this type. 
Specifically, the circuit in Fig.~\ref{Fig:circuit}  is described by 
 \begin{equation}\label{eq:h_k}
 \begin{split}
 d_x &= \ii\gamma - \kappa_1 +\kappa\cos{k_y} - t_x\cos{k_x},  \\
 d_y &= t_y\sin{k_y}-\kappa_2\sin{k_x}   
 \end{split} 
 \end{equation}
%
with a staggered on-site potential $d_z=\Delta$. 
The relation to circuit element parameters is 
$\gamma=\frac{1}{R},\kappa_1=\frac{1}{\omega L^0},\kappa=t_y=\omega C^0,t_x=\frac{1}{\omega}(\frac{1}{L^1}+\frac{1}{L^2}),\kappa_2=\frac{1}{\omega}(\frac{1}{L^2}-\frac{1}{L^1}),\Delta=\frac{1}{2\omega}(\frac{1}{L^A}-\frac{1}{L^B})\mycomment{-\omega(C^A-C^B)}$. Gain from the negative impedance converter (NIC) compensates for the loss incurred in the resistor $R$ enabling unimpeded signal propagation through the array \cite{Chen2009,Lee2003,Popa2012,Hofmann2019}. Dissipation for nH effect and spatially nonuniform elements for pMF can make $d_0$ respectively complex and inhomogeneous as per the mapping. As explained in SM the NIC element acting as a static negative resistor, together with other tunable grounded elements, guarantees a real-valued and uniform $d_0$ in analogy to a controllable chemical potential. 

\begin{figure}[t]
\centering
\includegraphics[width=0.4\textwidth]{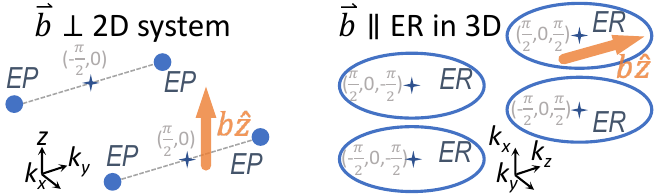}
\caption{Exceptional degeneracies in 2D and 3D circuit models generated by nH terms. Dashed bulk Fermi arc connects two EPs. Real energy spectrum emerges when pMF $\bm{b}$ takes the direction noted. In circuit calculations, we fix $\bm{b}=b\hat{z}$ for one exceptional degeneracy region as indicated.
}\label{Fig:cartoon}
\end{figure}

The circuit in Fig.~\ref{Fig:circuit} realizes relativistic band structures similar to graphene in a simpler square lattice. When $\gamma=\Delta=\kappa_2=0$ and $\kappa=\kappa_1$, the Hermitian spectrum $ E_{\bm{k}}$ of $h(\bm{k})$ without pMF, 
exhibits a pair of Dirac points located at $(\pm\frac{\pi}{2},0)$
with Fermi velocity $v_F^i=t_i$ in $i$-direction. 
As illustrated in Fig.~\ref{Fig:cartoon} inclusion of the dissipative term $\ii\gamma\sigma_x$ with $\gamma>\Delta$ splits  each Dirac point into a pair of EPs at $(\pm\frac{\pi}{2},\pm\sqrt{\gamma^2-\Delta^2}/v_F^y)$ at low energy.
Each EP pair is connected by a bulk Fermi arc, indicated by dashed lines in Fig.~\ref{Fig:cartoon}, along which the real part $\Re E_{\bm{k}}$ of the two bands touches \cite{Kozii2017,Papaj2019\mycomment{,Zhou2018,Carlstroem2018}} (see also Sec.~III in SM). We also consider in SM a 3D cubic noncentrosymmetric model of four parallel Weyl ERs depicted in Fig.~\ref{Fig:cartoon} and constructed by layering the 2D EP circuit along the $z$-axis with interlayer connections determined by $d_z=-t_z\cos{k_z}\mycomment{-\kappa_3}$.

A special feature of relativistic dispersions lies in that spatially varying hopping amplitudes can act as 
vector potentials chirally coupled to the low-energy excitations \cite{Guinea2009,Levy2010,Cortijo2015,Pikulin2016}. This feature naturally extends to the exceptional degeneracies when nH terms are included. A linear variation along the $x$-direction in inductances of the red elements in Fig.~\ref{Fig:circuit} produces spatial variation of the Hamiltonian parameter $\kappa_2=v_F^ybx$ and dictates open boundary along $x$. 
In the low-energy theory this manifests as a Landau gauge $A_y=bx$ giving rise to a uniform pMF $b\hat{z}$. Vector potential $A_x=-by$ can also be realized by varying $\kappa_1=\kappa-v_F^xb y$ along the $y$-direction with open edges
, see Fig.~S1\mycomment{\ref{Fig:circuit2}} in SM.
%

\begin{figure}[hbt]
\centering
\includegraphics[width=0.4\textwidth]{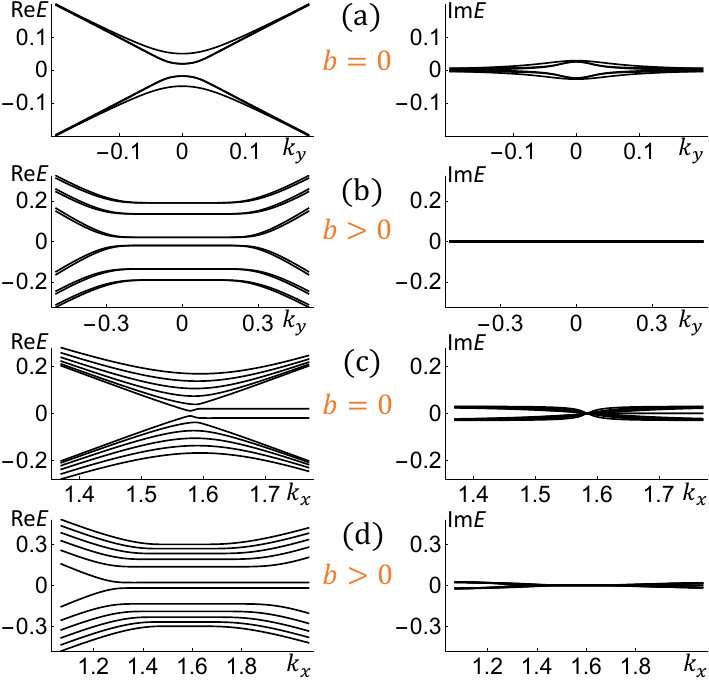}
\caption{Band structure of the $100\times 100$ square lattice EP circuit. 
Panels ab) have open boundary along $x$ realizing armchair-like bands under gauge $A_y=bx$ 
while panels cd) have open boundary along $y$ realizing zigzag-like bands under gauge $A_x=-by$. 
3D Weyl ER case at $k_z=\frac{\pi}{2}+\Delta/t_z$ is the same as cd). (See SM for extended discussion of the $b=0$ case.) We set $t_i=1$, unmodulated $\kappa=\kappa_1=1,\kappa_2=0$, and $\Delta=0.02,\gamma=0.03,b=0.009$. \mycomment{Each circuit direction has 100 sites.}}\label{Fig:bands}
\end{figure}

\textit{nH exceptional Landau levels} --
Band structure of the EP circuit is displayed in Fig.~\ref{Fig:bands}, where exceptional degeneracies are eliminated by the pMF illustrated in Fig.~\ref{Fig:cartoon}. 
Surprisingly, the resulting Landau-level-like flat bands exhibit spectra with consistently \textit{vanishing} imaginary part around the exceptional degeneracies. To understand this remarkable feature we first develop a low-energy theory of this nH Landau quantization of exceptional degeneracy and then discuss the origin of purely real spectrum. 

The low-energy Hamiltonian around the exceptional region with the magnetic field  $b\hat{z}$ indicated in Fig.~\ref{Fig:cartoon} reads 
\begin{equation}\label{h4}
h=\sum_{i=x,y}{(v_F^i\Pi_i+\ii\gamma_i)\sigma_i}+\Delta\sigma_z = \begin{bmatrix}
\Delta &  E_bf_+ \\
E_bf_- & -\Delta
\end{bmatrix}    
\end{equation}
where $v_F^i,b>0$, $E_b=\sqrt{2v_F^xv_F^yb}$ and $\Pi_i=p_i-A_i$. Momentum operator $p_i=-\ii\partial_i$ may be replaced by $k_i$ along the periodic direction of the circuit. $\Delta$ is replaced by $v_F^zk_z$ for the 3D ER case. Furthermore
\begin{equation}\label{h5}
    f_{\mp}=(v_F^x\Pi_x\pm\ii v_F^y\Pi_y+\ii\gamma_x\mp\gamma_y)/E_b.
\end{equation}
The first observation is that $[f_-,f_+]=1$ formally holds, even though $f_-^\dag\neq f_+$. Second, if $f_-\phi_0=0$ has a physical square-integrable solution, one can construct a tower of nH Landau levels (LL) through the wavefunction ansatz $\psi_n=(\alpha\phi_{n},\beta\phi_{n-1})^T$ for $n=0,1,2,\cdots$ where $\phi_{-1}=0$ and $\phi_{n>0}$ is obtained by the relation $f_+\phi_n=\sqrt{n+1}\phi_{n+1}\,,f_-\phi_n=\sqrt{n}\phi_{n-1}$. An explicit calculation then gives energy of the $n$th nH LL (LL$_n$) $E_{\mathrm{LL}_{n\pm}} = \pm\sqrt{\Delta^2+nE_b^2}$ when $n\geq1$ and $E_{\mathrm{LL}_{0+}} = \Delta$ when $n=0$, which is isospectral to the Hermitian counterpart. The construction is valid for arbitrary $\gamma_x,\gamma_y$ but breaks down in the presence of nonzero $\ii \gamma_z\sigma_z$. As this tilts the ER plane, requiring $\mathrm{pMF} \parallel\mathrm{ER}$ is thus the major difference in the 3D case in Fig.~\ref{Fig:cartoon}.

For the above procedure to work it is essential to ensure a physical solution of $f_-\phi_0=0$, the existence of which is not guaranteed in the nH case. Imagine for instance a spatially linear modulation of the resistors in the circuit, which can introduce an \textit{imaginary}-valued vector potential, e.g., $A_x=-\ii by$ and hence $[f_-,f_+]=\ii$ with a tower of complex LLs. In this case, however, a normalizable bounded solution of $f_-\phi_0=0$ does not exist \footnote{Avoiding this, one can instead use\blue{ nonuniformly varying} imaginary vector potentials that \mycomment{fully}decay at the boundary. 
}. 
We proceed as an example with our real-valued $A_x=-by$ that has a valid normalized wavefunction
\begin{equation}\label{eq:phi_n}
\phi_n
\mycomment{=(\sqrt{\pi}l_b n! 2^{n})^{-\frac{1}{2}}  \ee^{-y^2/2l_b^2 -(k_x+\ii\gamma_x-\gamma_y)y} H_n((y+\frac{k_x+\ii\gamma_x}{b})/l_b) }
=(\sqrt{\pi}l_b n! 2^{n})^{-\frac{1}{2}}  \ee^{-(y-y_0)^2/2l_b^2+\gamma_yy} H_n((y-y_0)/l_b)
\end{equation}
where $y_0=-(k_x+\ii\gamma_x)/b$, magnetic length $l_b=b^{-1/2}$ and $H_n(\mathsf{z})$ is the Hermite polynomial valued in the complex $\mathsf{z}$-plane. Note that $\gamma_x$ renders $\phi_n$ complex-valued  while $\gamma_y$ breaks its symmetry with respect to the Hermitian oscillation center $y=-k_x/b$. Henceforth we mainly consider the case with $\gamma_y=0$ which emerges naturally from our circuit realization. In general, nH systems have the potential for skin effect, which deviates from the Bloch band theory and the conventional bulk-boundary correspondence \cite{Lee2016,Alvarez2018a,Xiong2018,Kunst2018,Yao2018a,Yokomizo2019}. Remarkably here, not only is any possible skin aggregation suppressed in the low-energy regime as dictated by the \textit{bulk} magnetic confinement around $y_0$ at the length scale $\sqrt{n+1/2}l_b$, but also the nH quasiparticle is now oscillating along a \textit{complex} $y$-direction line centered at $y_0$. Therefore, this nH system under magnetic field indicates a new way to avoid the skin effect and provides a concrete example of a quasiparticle moving in the complex domain. This latter scenario is justified by the Hermite function actually being holomorphic on $\mathbb{C}$, although it is usually viewed solely as a real function in conventional quantum problems.  
The orthonormality, $\int_{-\infty}^\infty{\dd y \phi_n(\mathsf{z})\phi_m(\mathsf{z})} 
\mycomment{= (\sqrt{\pi}l_b n! 2^{n})^{-1}\int{\dd \mathsf{z} \ee^{-\mathsf{z}^2/2l_b^2} H_n(\mathsf{z})H_m(\mathsf{z})}}
= \delta_{mn}$ with $\mathsf{z}=(y-y_0)/l_b$, follows from analytic continuation. 
This way, one can also interpret the problem as analytically continuing the particle motion to the complex domain.

\textit{Spectral properties} --
Our nH low-energy theory has a real spectrum under finite pMF although it is not $\cal{PT}$-symmetric. To understand this one can formalize the above physical interpretation of a quasiparticle moving in the complex plane by defining an operator $\rho=\mathrm{diag}{(\ee^{\bm{\epsilon}\cdot\bm{p}},\ee^{\bm{\epsilon}\cdot\bm{p}})}$ 
that translates the system in real space along the imaginary direction by $\bm{\epsilon}=\frac{1}{b}\hat{z}\times\bm{\gamma}$ for $\bm\gamma=\gamma_x\hat x$. 
The pseudo-Hermiticity \cite{Mostafazadeh2003,Mostafazadeh2010}, a necessary but \textit{not sufficient} condition for a real spectrum, $\eta h \eta^{-1} = h^\dag$, holds here via a positive semi-definite Hermitian automorphism $\eta=\rho^\dag\rho$. In addition one can deduce the spectral reality via a similarity transformation $\rho h \rho^{-1}=h_0=h(\bm{\gamma}=0)$
%
%
which in general preserves the spectrum and maps $h$ 
to a Hermitian Hamiltonian with a spectral expansion $h_0=\sum_n E_n\ket{\varphi_n}\bra{\varphi_n}$ of real-spectrum conventional LLs. Then the left and right eigenstates, corresponding to Eq.~\eqref{eq:phi_n}, respectively of $h^\dag$ and $h$ are given by 
$\ket{\psi_n^{L(R)}}=\rho^{-1(\dag)}\ket{\varphi_n}$. 
Hence the biorthogonal representation \cite{Moiseyev2009,Brody2013}, $h=\sum_n E_n\ket{\psi_n^L}\bra{\psi_n^R}$, naturally follows. The aforementioned orthonormality based on nH Hermite functions helps prove herein the general pseudo-Hermitian orthonormality and biorthonormality $\braket{\psi_m^L|\eta|\psi_n^L}=\braket{\psi_m^R|\psi_n^L}=\delta_{mn}$.

Physically, the magnetic field in a relativistic system is crucial to the above reasoning. The phenomenon can be viewed as cancelling $\bm{\gamma}$ by absorbing it into the vector potential $\bm{A}$ in the kinetic term. This relies on $\bm{A}$ depending linearly on the spatial coordinate, necessary to give a uniform $\bm{b}$ field. 
One may wonder about the dual picture of translating by $\bm \gamma$ in the imaginary direction of the momentum space by using $\ee^{\bm\gamma\cdot\bm x}$ in $\rho$ with position operator $\bm x$, which actually explains under the gauge used the $\ee^{\gamma_yy}$ factor in Eq.~\eqref{eq:phi_n} by setting $\bm\gamma=\gamma_y\hat y$. It also relies on a finite $\bm b$, otherwise the wavefunction $\ket{\psi_n^{L(R)}}$ is unbounded. 
Therefore, magnetic field imparts a nonperturbative change to the system. The phenomenon and interpretation applies as well to the symmetric gauge, which we employ to construct the nH ground state and coherent state in SM.

\begin{figure}[hbt]
\includegraphics[width=0.4\textwidth]{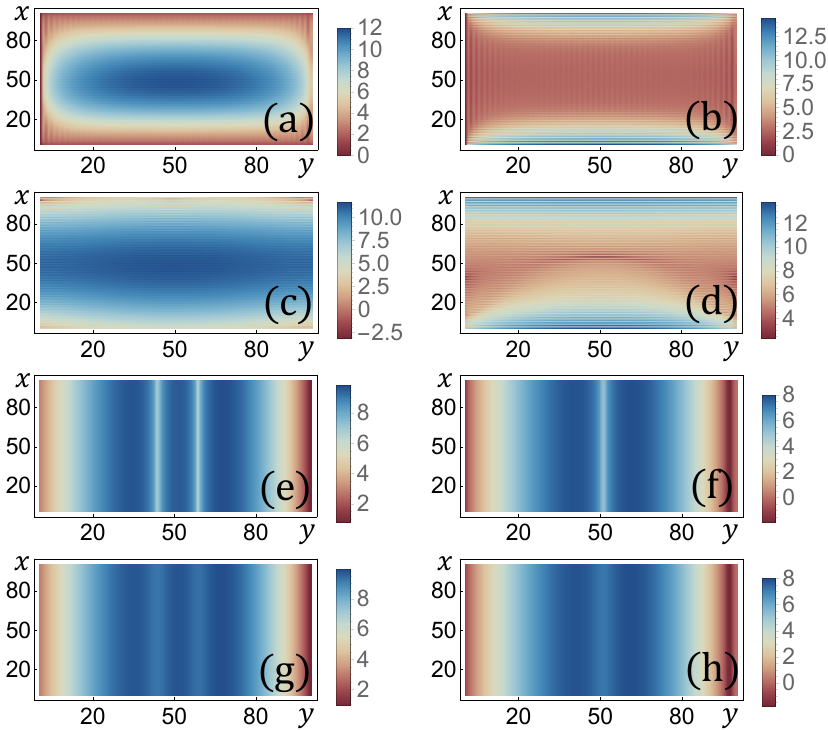}
\caption{Impedance and voltage tomography for the nH circuit used in Fig.~\ref{Fig:bands}. Amplitudes are plotted on logarithmic scale for armchair-like (abcd) and zigzag-like (efgh) cases. Panels abcd) show two cases with strongly enhanced response for LL$_{0\pm}$ in resonance. Impedance scan a) $Z_{(x,0),(x,y)}^{AA}(E_{\mathrm{LL}_{0+}})$, b) $Z_{(x,0),(x,y)}^{BB}(E_{\mathrm{LL}_{0-}})$ and voltage scan c) $V_{(L_x/2,L_y/2),(x,y)}^{AA}(E_{\mathrm{LL}_{0+}})$, d) $V_{(0,L_y/2),(x,y)}^{BB}(E_{\mathrm{LL}_{0-}})$. Panels efgh) show voltage scan displaying nH nodeful-nodeless transition of the up ($\phi_2$) and down ($\phi_1$) wavefuntion component of the LL$_{2+}$ in resonance. e) Hermitian, g) nH $V_{(x,y)}^{AA}(E_{\textrm{LL}_{2+}})$ and f) Hermitian, h) nH $V_{(x,y)}^{BB}(E_{\textrm{LL}_{2+}})$. \mycomment{Parameters are same as Fig.~\ref{Fig:bands}.}}\label{Fig:detection}
\end{figure} 
\textit{Detection schemes} --
Based on the KCL construction, one can readily predict the directly measurable electrical response of the circuit. We consider two types of circuit tomography assuming system size $L_x\times L_y$, (i) impedance scan $Z_{(x,0),(x,y)}^{s_0s}(d_0)$ reflecting a direct impedance measurement between two points $(x,0,s_0)$ and $(x,y,s)$, and (ii) voltage scan $V_{(x_0,y_0),(x,y)}^{s_0s}(d_0)$ probing voltage at any node $(x,y,s)$ in response to a current input at the midpoint $y_0=L_y/2$ of the $x_0=0,L_x/2$ lines, where $s,s_0=A,B$. We derive in SM expressions for both quantities in terms of left and right eigenstates of the nH Hamiltonian. 

Several observations can be made based on the predictions for impedance and voltage scans in Fig.~\ref{Fig:detection}. First, in order to have a significant voltage response, large density of states within a small range of admittance eigenvalue $j$ is required. Compared to topological boundary zero modes \cite{Lee2018}, this is naturally achieved in the presence of pMF by the flat nH LLs, which can be set in resonance by controlling $d_0$. An example of this enhancement is given in Fig.~\ref{Fig:detection}abcd. 
Second, a unique sublattice polarization of the lowest LL (LL$_0$) and the general wavefunction form $\psi_n=(\alpha\phi_{n},\beta\phi_{n-1})^T$ hold for the exceptional LLs. Controlling $d_0,s_0,s$, sublattice-resolved responses provide access to this. 
The armchair-like case has every nH LL doubly degenerate in a pMF while the zigzag-like case mixes the nH LL$_0$s with the edge states. Below we use both to highlight different features. 

\textit{Edge state from nH energy-reflection symmetry} --
Consider the EP circuit in resonance at $d_0=E_{\mathrm{LL}_{0\pm}}=\pm\Delta$, i.e., the positive/negative LL$_{0\pm}$ in Fig~\ref{Fig:bands}b. One thus has dichotomous choices in $d_0,s_0,s$ and $x_0=0,L_x/2$. 
Fig.~\ref{Fig:detection}ac and \ref{Fig:detection}bd illustrate the \textit{only} two enhanced cases respectively of bulk and edge nature as seen from the pronounced signal distribution contrast. All others are largely suppressed or vanishing. The edge state LL$_{0-}$ localized around $x_0=0,L_x$, surprisingly, cannot be captured in a low-energy nH $2$-flavour $2$D massive Dirac theory under pMF, which solely leads to two degenerate $\mathrm{LL}_{0+}$ states. Analysed in SM, it is actually the consequence of a strong lattice nH energy-reflection symmetry for \textit{any} Hermitian or nH bipartite hoppings which is beyond the usually pertinent chiral or particle-hole symmetry. 
These confirm the nH sublattice polarization from an intricate interplay between the pMF, Dirac mass, armchair-like bands, and the nH symmetry that dictates pairs of opposite and however complex or real bands.

\textit{nH nodeless wavefunction} --
Observation of the non-Hermiticity is most prominent via inspecting the wavefunctions because of the spectral property discussed. The two-component general wavefunction form here becomes relevant. One can combine the nodal structure of conventional Hermite functions, i.e., $H_n(y)$ possesses $n$ nodes, with our physical interpretation of translating the motion to the complex plane of $H_n(\mathsf{z})$. This directly leads to the removal of all nodes by the finite $\Im \mathsf{z}=\epsilon$. Therefore, a transition from nodeful to \textit{nodeless} probability (voltage) distribution becomes a distinguishing nH feature. This is made practically feasible by the quantum superposition principle, i.e., one can inject spatially sinusoidally oscillating current at 
a certain wavenumber $k_x'$ along one single open boundary, say, the $y=0$ edge of the 
EP \mycomment{(ER) }circuit, which suffices to extract the nH Hermite wavefunction associated with 
$k_x'$. 
Fig.~\ref{Fig:detection}eg and \ref{Fig:detection}fh exemplify this nodeful-nodeless transition of $\phi_2$ and $\phi_1$ respectively by plotting the amplitude of voltage response. 


\textit{Outlook} --
Using specially designed ac electric circuits we develop a theory and present detection schemes for a unique nH low-energy real spectrum without skin effect, which arise from the relativistic exceptional degeneracies under magnetic field and exhibit nH symmetry protected edge state and quasiparticle moving in the complex domain. These results enrich a novel platform for synthetic quantum systems and lay the groundwork for future investigations of the interplay between non-Hermiticity and magnetic field, which is relevant to the emerging real quantum systems with exceptional degeneracies \cite{Dembowski2001,Dembowski2004,Gao2015,Zhen2015,Cerjan2019}. Various intriguing questions are to be explored ahead, including imaginary-valued vector potential or magnetic field, further generalization of quasiparticle living in the complex plane\mycomment{ using other mathematical functions}, nH quantum valley Hall effect in the EP circuit with pMF, and a possible nH Hofstadter butterfly readily realized by introducing resistors to the circuit of nodes with internal eigenmodes in a similar manner to the present study \cite{Albert2015,Zhao2018,SM}.

\let\oldaddcontentsline\addcontentsline
\renewcommand{\addcontentsline}[3]{}

\begin{acknowledgments}
X.-X.Z appreciates discussions with R. Haenel, W. Yang, \'{E}. Lantagne-Hurtubise, and T. Liu. Research described in this article was supported by NSERC and by CIfAR.
\end{acknowledgments}\mycomment{\Yinyang}

\bibliography{reference.bib}  
\let\addcontentsline\oldaddcontentsline

\newpage
\onecolumngrid
\newpage
{
	\center \bf \large 
	Supplemental Material\\
	\large for ``\newtitle"\vspace*{0.1cm}\\ 
	\vspace*{0.5cm}
}
\twocolumngrid	

\tableofcontents


\setcounter{equation}{0}
\setcounter{figure}{0}
\setcounter{table}{0}
\setcounter{page}{1}
\renewcommand{\theequation}{S\arabic{equation}}
\renewcommand{\thefigure}{S\arabic{figure}}
\renewcommand{\theHtable}{Supplement.\thetable}
\renewcommand{\theHfigure}{Supplement.\thefigure}
\renewcommand{\bibnumfmt}[1]{[S#1]}
\renewcommand{\citenumfont}[1]{S#1}

\section{Model Hamiltonians}\label{App:models}

A simple formulation largely owes to the specially tuned positions of Dirac or Weyl points and the form of the model. In fact, various other graphene models with pMF, including the original honeycomb one, in general lead to imperfect low-energy behaviour (armchair) and especially unflat LLs (zigzag) that will be reported elsewhere \cite{Lantagne-Hurtubise2019}. 
These demerits for detection can be overcome by the following minimal square and cubic lattice models. 

The general Hamiltonian of the square lattice graphene-like EP circuit with pMF is 
%
\begin{equation}\label{eq:H_sq_graphene}
\begin{split}
&H= \sum_r  \ii\gamma (a_r^\dag b_r + b_r^\dag a_r)  + \Delta (a_r^\dag a_r - b_r^\dag b_r) \\
& + \{ a_r^\dag [ -2\kappa_1b_r + (\kappa-t_y) b_{r+\hat{y}}+(\kappa+t_y)b_{r-\hat{y}} 
\mycomment{- \frac{t}{2} (b_{r+\hat{y}}-b_{r-\hat{y}})}  \\
&  - (t_x-\kappa_2) b_{r+\hat{x}}-(t_x+\kappa_2)b_{r-\hat{x}} ] /2 + \mathrm{H.c.} \}  .
\end{split} 
\end{equation}
The cubic lattice noncentrosymmetric Weyl ER circuit can be obtained by replacing the $\Delta$-term in Eq.~\eqref{eq:H_sq_graphene} by $- [ \kappa_3a_r^\dag a_{r} + t_z ( a_r^\dag a_{r+\hat{z}} + a_r^\dag a_{r-\hat{z}})] + [a\rightarrow b]$. 
Resistors connecting $A,B$ nodes in general generate off-diagonal nH terms while unequal resistors grounding inequivalent nodes generate nH $\sigma_z$ terms. Without loss of generality, we mainly focus on $\ii\gamma\sigma_x$ by resistively connecting $A,B$ nodes in a primitive cell.  
For the 2D case, as long as $\Delta=0$, the system exhibits two Dirac points since the inversion symmetry $h(\bm k)=\sigma_xh(-\bm k)\sigma_x$ 
remains. 
Let us state the more general Weyl case that obviously reduces to the graphene-like case by dropping the $z$-direction dependence. When $\kappa=\kappa_1,\kappa_2=\kappa_3=\gamma=0$, the Fermi velocity at four nodes $(m_x\frac{\pi}{2},0,m_z\frac{\pi}{2})$ is $v_F=(m_xt_x,t_y,m_zt_z)$ for $m_{x/z}=\pm$. 
The finite nH $\gamma$ term 
brings the 3D Weyl band to an elliptical ER around the original Weyl point with semi-major and semi-minor axes $\gamma/|v_F^{y,z}|$ in the $k_y,k_z$-plane as shown in Fig.~\ref{Fig:cartoon}. 

Let us focus on the $(\frac{\pi}{2},0,\frac{\pi}{2})$ node, around which a vector potential $\bm A$ satisfying $\bm v_F\cdot\bm A = (\kappa_1-\kappa,\kappa_2,\kappa_3)$ is introduced by uniformly modulating $\kappa_i$'s in a circuit accordingly where we assume a positive elementary charge set to unity. This can be seen by expanding the $k$-space Hamiltonian $h(\bm k)$ around the $(\frac{\pi}{2},0,\frac{\pi}{2})$ node up to the leading order, 
\begin{equation}
\begin{split}
d_x&=t_x(k_x-\frac{\pi}{2})+\kappa-\kappa_1+O[(k_x-\frac{\pi}{2})^2]\\
d_y&=t_y k_y-\kappa_2+O(k_y^2)\\
d_z&=t_z(k_z-\frac{\pi}{2})-\kappa_3+O[(k_z-\frac{\pi}{2})^2],    
\end{split}
\end{equation}
which is justified by assuming $\kappa_i$'s modulation is spatially slow and thus mismatched in Fourier space with the lattice. For instance, a pMF $b\hat z$ is generated by making $\kappa_1=\kappa-v_F^xb y$ ($\kappa_2=v_F^ybx$) that amounts to a $y$ open ($x$ open) geometry with Landau gauge $A_x=-by$ ($A_y=bx$), which applies to the 2D EP circuit as well. Particular to the Weyl ER case, we can also use $\kappa_3=-v_F^zbx$ ($\kappa_3=v_F^zby$) to generate pMF $b\hat y$ ($b\hat x$).

The Hamiltonian of another noncentrosymmetric cubic lattice Weyl ER circuit is 
\begin{equation}
\begin{split}
d_x &= \ii\gamma - \kappa_1+\kappa\cos{k_x}+\sin{k_x}\sin{k_z}\\
d_y &= \sin{k_y}-\kappa_2\sin{k_z}\csc^3{k_R}(\cos{k_R}\cos{k_z}-\cos{2k_R})\\
d_z &= -(\cos{k_z}-\cos{k_L})(\cos{k_z}-\cos{k_R})\\
&-2(2-\cos{k_x}-\cos{k_y}).    
\end{split}
\end{equation}
It reads in the real space 
\begin{equation}
\begin{split}\label{eq:Hsymm}
&H= \sum_r   \mycomment{\ii\gamma a_r^\dag b_r} \ii\gamma(a_r^\dag b_r + b_r^\dag a_r) 
+ \{ a_r^\dag [ - t_{2z} (a_{r+2\hat{z}}+a_{r-2\hat{z}})  \\
&  -t_0 a_r + \sum_{i=x,y,z}{t_i (a_{r+\hat{i}}+a_{r-\hat{i}})}  ]  - (a\rightarrow b)  \}   \\
&+ \{ a_r^\dag [ g_y (b_{r-\hat{y}}-b_{r+\hat{y}}) - \kappa_1 b_r +\kappa/2 (b_{r+\hat{x}}+b_{r-\hat{x}})  \\
&  + g_{xz} (b_{r+\hat{x}-\hat{z}}+b_{r-\hat{x}+\hat{z}} - b_{r+\hat{x}+\hat{z}}-b_{r-\hat{x}-\hat{z}})   \\
& - g_{z} (b_{r+\hat{z}}-b_{r-\hat{z}}) + g_{2z} (b_{r+2\hat{z}}-b_{r-2\hat{z}})] + \mathrm{H.c.} \}  ,
\end{split} 
\end{equation} 
in which $t_0=\cos{k_L}\cos{k_R}+\frac{9}{2}$, $t_x=t_y=1$, $t_z=\frac{1}{2}(\cos{k_L}+\cos{k_R})$, $g_y=\frac{1}{2}$, $g_{z}=\frac{\kappa_2}{2}\csc^3{k_R}\cos{2k_r}$, $g_{2z}=\frac{\kappa_2}{4} \csc^3{k_R}\cos{k_R}$, and $g_{xz}=t_{2z}=\frac{1}{4}$. It is generalized from a previous noncentrosymmetric cubic lattice Weyl semimetal model with the minimal four Weyl points \cite{XXZ:Luttinger}. When $\gamma=0,\kappa=\kappa_1=1,\kappa_2=0$, the anisotropic Fermi velocities $\bm v_F$ of the four Weyl points at $(0,0,-k_R)$, $(0,0,-k_L)$, $(0,0,k_L)$, $(0,0,k_R)$ are respectively
$(-\sin k_R,1,c_{LR}\sin k_R )$, $(-\sin k_L,1,-c_{LR}\sin k_L )$, $(\sin k_L,1,c_{LR}\sin k_L )$, $(\sin k_R,1,-c_{LR}\sin k_R )$, wherein $c_{LR}=\cos k_L-\cos k_R$. When $\kappa_1=\kappa-v_F^xb y/2$ and $\kappa_2=v_F^ybx/2$, it exhibits a symmetric gauge $\bm A=(-by,bx,0)/2$ of pMF $b\hat{z}$ for the $k_R$ node or the associated ER and the like for Landau gauges. 

Here, the special form is necessary to validate $A_y$ since both the pseudospin $\sigma_y$ and the modulation term in $d_y$ must be odd under time reversal $\mathcal{T}$. 
This model is more complex but useful in that four position-tunable Weyl nodes or ERs are located or centered on the \textit{same line}, the $z$-axis. Therefore, there will be no node or ER superposition in circuit calculations when one sets open boundary condition in both $x$ and $y$, which is used to show the symmetric gauge wavefunction in Sec.~\ref{App:gauge}. 


\section{Circuit construction}
\subsection{Mapping RLC circuits onto Bloch Hamiltonians}
\begin{figure}[hbt]
\centering
\includegraphics[width=0.33\textwidth]{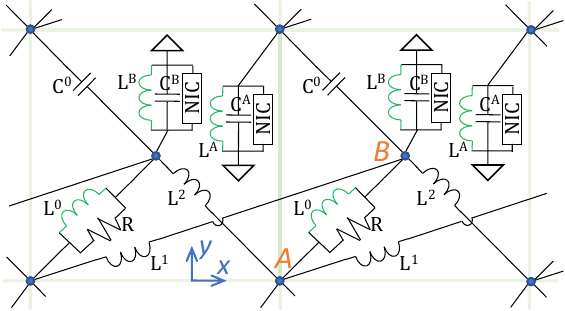}
\caption{Circuit for square lattice \mycomment{(pale green) }graphene-like EP model with Landau gauge $A_x=-by$ and pMF $b\hat{z}$ generated by the green elements modulated along $y$-direction. Open (periodic) boundary condition in $y$ ($x$). 
}\label{Fig:circuit2}
\end{figure}
\begin{widetext}
With the Kirchhoff current law and voltage law, one can apply either the node analysis or the loop analysis to a circuit. In a Lagrangian formulation, these can be expressed in the phase space of node flux variable $\varphi_i$ and loop charge variable $q_i$, respectively. The node analysis employed in this study is useful for identifying a quantum lattice analog Hamiltonian since the number of independent KCL equations matches exactly the number of lattice unit cells $N$ if one introduces the auxiliary ground node.

Let us explicitly work out the mapping between the ac electric circuit indicated in Fig.~\ref{Fig:circuit} and the 2D EP model Hamiltonian Eq.~\eqref{eq:h_k} or Eq.~\eqref{eq:H_sq_graphene} as an example. We also consider here a slightly more general circuit depicted in Fig.~\ref{Fig:circuit2}. The two circuits share the same basic structure but differ in the composition of the grounded elements and the gauge potential realized. The general Lagrangian including both cases reads
\begin{equation}
    L=\frac{1}{2}\sum_r{[ C^0(\dot\varphi_r^A - \dot\varphi_{r-\hat y}^B)^2 -\frac{1}{L_r^0}(\varphi_r^A - \varphi_{r}^B)^2 -\frac{1}{L_{r}^1}(\varphi_r^A - \varphi_{r+\hat x}^B)^2 -\frac{1}{L_{r}^2}(\varphi_r^A - \varphi_{r-\hat x}^B)^2 
    + \sum_{\alpha=A,B}{(C^\alpha {\dot{\varphi}_r^{\alpha\,2}} - \frac{1}{L_r^\alpha} {\varphi_r^{\alpha\,2}})}]
    }
\end{equation}
with the Rayleigh dissipation function
\begin{equation}\label{eq:D}
    D=\frac{1}{2}\sum_r{ \frac{1}{R}(\dot\varphi_r^A - \dot\varphi_{r}^B)^2 - \frac{1}{R}({\dot\varphi_r^{A\,2}} + {\dot\varphi_r^{B\,2}})}.
\end{equation}
Note that for completeness we attach space coordinate index $r$ to all the colored elements in both Fig.~\ref{Fig:circuit} and Fig.~\ref{Fig:circuit2}, 
to indicate their potential spatial modulation required for the generation of the pMF. The second term in $D$ represents the NIC element realizing a static negative resistor with resistance $-R$. Using the Euler-Lagrange Eq.~\eqref{eq:EL_eq} for $\varphi_r^A$ and $\varphi_r^B$, we obtain the KCL equations 
\begin{equation}\label{eq:J_rspace}
\begin{split}
    &\ii \left\{ \left[-\omega(C^0+C\mycomment{_r}^A)+\frac{1}{\omega}(\frac{1}{L_r^0}+\frac{1}{L_r^1}+\frac{1}{L_r^2}+\frac{1}{L_r^A})\right]v_r^A 
    + (-\frac{1}{\omega L_r^0}+\frac{\ii}{R})v_r^B - \frac{1}{\omega L_r^1}v_{r+\hat x}^B - \frac{1}{\omega L_r^2}v_{r-\hat x}^B + \omega C^0 v_{r-\hat y}^B \right\} = i_r^A \\
    &\ii \left\{  
    (-\frac{1}{\omega L_r^0}+\frac{\ii}{R})v_r^A - \frac{1}{\omega L_{r-\hat x}^1}v_{r-\hat x}^A - \frac{1}{\omega L_{r+\hat x}^2}v_{r+\hat x}^A + \omega C^0 v_{r+\hat y}^A
    + \left[-\omega(C^0+C\mycomment{_r}^B)+\frac{1}{\omega}(\frac{1}{L_r^0}+\frac{1}{L_{r-\hat x}^1}+\frac{1}{L_{r+\hat x}^2}+\frac{1}{L_r^B})\right]v_r^B
    \right\} = i_r^B. 
\end{split}
\end{equation}
The complete set of such equations at all space positions $r$ constitutes the desired form of the admittance problem $J \bm{v}= \bm{i}$. To see this, leaving out the diagonal part (term in the square brackets), one immediately identifies the mapping $J=\ii H$ with hopping amplitudes $\omega C$ ($\frac{-1}{\omega L}$) for nodes connected by a capacitor (inductor) while a nH hopping $\ii /R$ accounts for a resistor. The general form of the diagonal term for node $n$ reads similarly but with an overall minus sign 
\begin{equation}\label{eq:Hnn}
H_{nn}= -\sum_{i}{\omega C_i} + \sum_j{\frac{1}{\omega L_j}}-\sum_k{\frac{\ii} {R_k}}    
\end{equation}
 where $i,j,k$ run respectively over all inequivalent $C,L,R$ elements connected to node $n$, including any extra elements grounding the node\mycomment{ \cite{Lee2018}}. Here $n$ stands for both $r$ and the sublattice $A,B$. Note that, owing to the inclusion of the grounded NIC element in the circuit we no longer have in Eq.~\eqref{eq:J_rspace} the imaginary part in Eq.~\eqref{eq:Hnn}.

To understand this circuit more clearly, we temporarily drop the spatial dependence of the elements and transform to the $k$-space representation
\begin{equation}
    \ii
    \begin{bmatrix}
    -\omega(C^0+C^A)+\frac{1}{\omega}(\frac{1}{L^0}+\frac{1}{L^1}+\frac{1}{L^2}+\frac{1}{L^A}) & \omega C^0 \ee^{-\ii\bm k\cdot\hat y} - \frac{1}{\omega} (\frac{1}{L^0} + \frac{1}{L^1}\ee^{\ii\bm k\cdot\hat x} + \frac{1}{L^2}\ee^{-\ii\bm k\cdot\hat x}) +  \frac{\ii}{R} \\
    \omega C^0 \ee^{\ii\bm k\cdot\hat y} - \frac{1}{\omega} (\frac{1}{L^0} + \frac{1}{L^1}\ee^{-\ii\bm k\cdot\hat x} + \frac{1}{L^2}\ee^{\ii\bm k\cdot\hat x}) +  \frac{\ii}{R} & -\omega(C^0+C^B)+\frac{1}{\omega}(\frac{1}{L^0}+\frac{1}{L^1}+\frac{1}{L^2}+\frac{1}{L^B})
    \end{bmatrix}
    \begin{bmatrix}
    v^A \\
    v^B 
    \end{bmatrix}
    =
    \begin{bmatrix}
    i^A \\
    i^B 
    \end{bmatrix}.
\end{equation}
This gives the two-band Hamiltonian $h(\vec{k}) = d_0\mathbb{1}+\sum_i{d_i\sigma_i}$ with 
\begin{equation}\label{eq:all_d}
\begin{split}
    d_x &= \frac{\ii}{R} - \frac{1}{\omega L^0} +\omega C^0\cos{k_y} - \frac{1}{\omega}(\frac{1}{L^1}+\frac{1}{L^2})\cos{k_x}\,, \qquad
    d_y = \omega C^0\sin{k_y}-\frac{1}{\omega}(\frac{1}{L^2}-\frac{1}{L^1})\sin{k_x}\,, \\
    d_z &=  \frac{1}{2} [\frac{1}{\omega}(\frac{1}{L^A}-\frac{1}{L^B})-\omega(C^A-C^B)] \,,\qquad
    d_0 = \frac{1}{\omega}(\frac{1}{L^0}+\frac{1}{L^1}+\frac{1}{L^2})-\omega C^0 + \frac{1}{2} [\frac{1}{\omega}(\frac{1}{L^A}+\frac{1}{L^B})-\omega(C^A+C^B)].
\end{split}
\end{equation}
Thus, we arrive at the mapping between the circuit parameters and parameters entering the Hamiltonian 
\begin{equation}
\gamma=\frac{1}{R},\kappa_1=\frac{1}{\omega L^0},\kappa=t_y=\omega C^0,t_x=\frac{1}{\omega}(\frac{1}{L^1}+\frac{1}{L^2}),\kappa_2=\frac{1}{\omega}(\frac{1}{L^2}-\frac{1}{L^1}),2\Delta=\frac{1}{\omega}(\frac{1}{L^A}-\frac{1}{L^B})-\omega(C^A-C^B).  \end{equation}
\end{widetext}

\subsection{Role of grounded elements}
As seen from Eq.~\eqref{eq:Hnn}, a direct lattice counterpart of the Hamiltonian  Eq.~\eqref{eq:h_k} or Eq.~\eqref{eq:H_sq_graphene} will possibly make the diagonal part $d_0$ of the circuit Hamiltonian complex and, in the presence of spatially varying elements, inhomogeneous. In the quantum description this corresponds to a complex-valued and spatially varying chemical potential which leads to various complications.  As we show below, however, \textit{grounded} elements can be designed to counteract this undesirable effect.

\begin{figure}[hb]
\includegraphics[width=0.5\columnwidth]{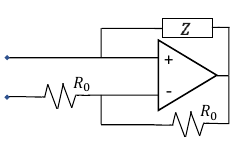}
\caption{Negative impedance converter circuit.}\label{Fig:NIC}
\end{figure}

First, to cancel the imaginary part of $H_{nn}$ caused by $R$ connecting $A,B$, one needs a matching static negative resistor $-R$ grounding $A,B$, which can be realized by the NIC element whose structure is depicted Fig.~\ref{Fig:NIC}. In our calculations we already included this contribution in Eq.~\eqref{eq:D}.
The NIC element, as is in landline repeaters and active filters, is routinely used in (integrated) analog circuits to cancel undesired lossy resistances \cite{Chen2009,Lee2003,Popa2012,Hofmann2019} and we briefly describe its basic principle of operation and properties.
It is well known in electrical engineering that 
a non-inverting operational amplifier of negative feedback connected with two equal resistors will effectively reverse the impedance $Z$ of a third element connecting the operational amplifier's output and the non-inverting input. This realizes a negative impedance converter with current inversion. Putting a desired resistor $R$ as the third element thus generates an effectively static negative resistor of $-R$. It works as a static linear circuit, operates over a broad range of voltages and frequencies, and is readily available from the LM741, LM324, MAX4014 amplifiers \cite{Chen2009,Lee2003,Popa2012\mycomment{,Hofmann2019}}.


Second, to compensate for the possible spatial inhomogeneity in $H_{nn}$ brought about by spatial variation of the ungrounded $L,C$ elements (required for generating pMF), one needs to spatially vary the impedance of the grounded elements in some cases. As shown in the main text, the Landau gauge $A_y=bx$ is generated by spatially modulating $\kappa_2$ while keeping $t_x$ fixed, i.e., modulating the values of $L^1$ and $L^2$  in an equal but opposite manner. In this case, $d_0$ by itself remains constant in space. However, the case of the other gauge potential, $A_x=-by$, generated by modulating $\kappa_1$ and hence $L^0$, is different. Eq.~\eqref{eq:all_d} implies that modulating $L^0$ will inevitably cause spatial inhomogeneity in $d_0$. This would interfere with our desired mapping between the circuit and the target Hamiltonian. The solution is to take advantage of the grounded elements $L^{A,B},C^{A,B}$. For instance, one can make $\frac{1}{L^A}+\frac{1}{L^B}$  vary to compensate for the variation in $\frac{1}{L^0}$ while keeping $\frac{1}{L^A}-\frac{1}{L^B}$ and thus $d_z$ fixed. This leads to a spatially uniform $d_0$. 

Finally, in either gauge, one can tune the global uniform value of $\frac{1}{L^A}+\frac{1}{L^B}$ to control the value of $d_0$. More generally, if $C^{A,B}$ are present as in the case of Fig.~\ref{Fig:circuit2}, one can fix $C^A = C^B$ and vary $C^A + C^B$ together with $\frac{1}{L^A}+\frac{1}{L^B}$. This is
analogous to tuning the value of the chemical potential in the analog quantum Hamiltonian.

\section{Energy spectrum along and across bulk Fermi arc without pMF}
We give here some details related to Fig.~\ref{Fig:bands}ac and the bulk Fermi arc. 

\begin{figure}[hbt]
\centering
\includegraphics[width=0.4\textwidth]{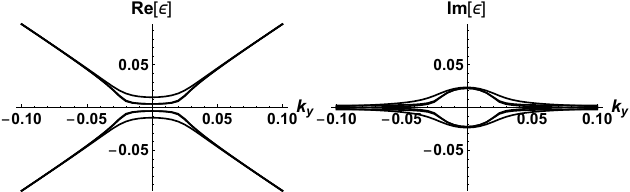}
\caption{Two EPs connected by a bulk Fermi arc for the square lattice EP model without pMF. Same plot as Fig.~\ref{Fig:bands}a except that each circuit direction is enlarged to 500 sites.
}\label{Fig:band_bigN}
\end{figure}

Between the two EPs at $(\frac{\pi}{2},\pm\sqrt{\gamma^2-\Delta^2}/v_F^y)$ when $\gamma>\Delta$, there is a bulk Fermi arc as illustrated in Fig.~\ref{Fig:cartoon}, where the real (imaginary) part of the spectrum is vanishing (finite). We present in Fig.~\ref{Fig:bands}ab the case of a practically more feasible size. It gains a sharper appearance when the circuit size expands, as shown in Fig.~\ref{Fig:band_bigN}. One can recognize the Fermi arc length roughly of the analytic value $2\sqrt{\gamma^2-\Delta^2}/v_F^y=0.045$.

In Fig.~\ref{Fig:bands}cd, we show the band structure along $k_x$ that aids best in presenting the flat bands although the ER is not directly visible in Fig.~\ref{Fig:bands}c if we regard it as the 3D case. The band structure can be still understood from the spectrum
\begin{equation}\label{eq:ER_cut}
    E=\pm\sqrt{(k_x+\ii\gamma)^2+k_y^2+k_z^2}
\end{equation}
although $k_y$ is no longer a good quantum number and instead it here takes quantized values as per the problem with open boundary in $y$. Given a $k_z$ smaller than the ER radius, Eq.~\eqref{eq:ER_cut} is exactly the spectrum of a pair of EPs along $k_y$ in 2D, where $\Re E (k_x)$ ($\Im E (k_x)$) has an apex cusp (a finite jump) at $k_x=0$ if $k_y$ crosses the bulk Fermi arc connecting two EPs. Therefore, Fig.~\ref{Fig:bands}c is an aggregation of cuts at different allowed $k_y$'s of two intersecting EPs between the ER and the given  $k_z\mycomment{=\frac{\pi}{2}+\Delta/t_z}$-plane, which explains the abrupt sign change across the bulk Fermi arc in the imaginary part.

\section{Energy spectrum with edge state and nH energy-reflection symmetry}
We first explain how the low-energy Dirac theory fails to capture the edge state and then present the correct nH symmetry to resolve this. 

Focusing on the low-energy property of the nH EP circuit, one can formulate a nH $2$-flavour $2$D massive Dirac theory under pMF, 
\begin{equation}\label{eq:2F_Dirac}
h_\mathrm{2F}=p_x\alpha_1+(p_y+\ii\gamma_y)\alpha_2+\Delta\alpha_0-(A_x-\ii\gamma_x)\beta_1-A_y\beta_2,
\end{equation}
to account for both the valley (flavour $\tau$) and sublattice ($\sigma$) pseudospins. Here, $\alpha_1=\sigma_x\otimes\tau_z,\alpha_2=\sigma_y\otimes\tau_0,\alpha_0=\sigma_z\otimes\tau_0,\beta_1=\sigma_x\otimes\tau_0,\beta_2=\sigma_y\otimes\tau_z$ and we also define  $\alpha_3=\sigma_x\otimes\tau_x,\alpha_4=\sigma_x\otimes\tau_y$ since $\alpha_{0,1,2,3,4}$ form a maximal set of mutually anticommuting $\Gamma$-matrices. 
The Dirac mass effectively opens $\pm\Delta$ gap at $K$ and $K'$ valleys of opposite chirality. The pMF chirally coupled to two valleys makes the nH LL$_0$s at two valleys \textit{both} localized to sublattice $A$ and shifted upwards to $E_\mathrm{LL_{0+}}=E_{\mathrm{LL}_0}=\Delta$ because of the $\mathcal{T}$-symmetry when $\bm\gamma=0$. Therefore, despite the coincidence of two valleys, we are able to observe this unique behavior in the armchair-like case and hence avoid the zigzag-like edge states mixing up with LL$_0$s. 

However, there is at all \textit{no} partner state of energy $E_\mathrm{LL_{0-}}=-E_{\mathrm{LL}_0}$, which is present in the circuit calculation as shown in Fig.~\ref{Fig:bands}b in the main text.
In fact, one cannot find an energy-reflection symmetry for $h_\mathrm{2F}$ unless in the absence of pMF.
When $\bm A\neq0$, momentum $\bm p$ and $\bm A$ do not commute and a unitary matrix anticommuting with $h_\mathrm{2F}$, i.e., an energy-reflection symmetry, cannot be found. 
To see this, note that $\{\alpha_{3},\beta_1\}\neq0,\{\alpha_{4},\beta_1\}\neq0$ are linearly independent and so is for $\beta_2$. Thus, any linear combination of $\alpha_{3,4}$, the sole possibility anticommuting with $\alpha_{0,1,2}$, will not anticommute with $\beta_{1,2}$. 
On the other hand, when $\bm A = 0$, pairs of opposite complex energies can be directly solved as momenta become good quantum numbers.

The paradox is resolved by a strong nH energy-reflection 
symmetry $\mathcal M$ in lattice systems with \textit{any} Hermitian or nH bipartite hoppings, staggered potential $\Delta$, and spatial modulation pattern for pMF. The Hamiltonian takes the form 
\begin{equation}
    H=\begin{bmatrix}
    \Delta I      & B \\
    C       & -\Delta I
\end{bmatrix}
\end{equation}
in the basis $(a_1^\dag,\dots,a_N^\dag,b_1,\dots,b_N)$ where $B,C$ are \textit{general} square matrices not necessarily related by Hermitian conjugate. It bears the nH energy-reflection symmetry as seen from the characteristic polynomial 
\begin{equation}
|H - E I| = |(E^2-\Delta^2)I - BC|,    
\end{equation}
i.e., (complex) eigenenergies $\pm E$ always come in pair. Two special cases are as follows. When $\Delta=0$, it gives a nH chiral symmetry 
\begin{equation}
\mathcal{S}H\mathcal{S}^{-1}=-H
\end{equation}
by $\mathcal S=\sigma_z$ in the sublattice space. In the EP models with real hoppings only, thought as a BdG Hamiltonian, it amounts to a nH particle-hole symmetry 
\begin{equation}
\mathcal{C}H\mathcal{C}^{-1}=-H
\end{equation}
by mapping $c_i\rightarrow c_i^\dag$ for $c=a,b$.
For the 3D Weyl ER case, non-bipartite hoppings due to the $d_z\sigma_z$-terms in general could break $\mathcal M$. However, as long as $d_z$ is a function of good-quantum-number momenta only (say, $k_i$, i.e., at least periodically connected along $i$-axis) and no modulation for generating pMF in $d_z$, the $k_i$-dependent system retains $\mathcal M$.

For the EP case, these symmetry properties as well hold when we set periodic boundary condition in $x$ ($y$) and work in the 1D\mycomment{ zigzag-like (armchair-like)} model dependent on $k_x$ ($k_y$). Therefore, $\cal{M}$ promises the appearance of $\pm E_k$ in the spectrum. From Fig.~\ref{Fig:detection}bd in the main text, we know that this LL$_{0-}$ state not captured in the low-energy theory is only polarized to sublattice $B$ and localized to the both edges. Along with the foregoing analysis of the LL$_{0+}$ state, this immediately determines that LL$_{0-}$ actually originates from the opposite-to-bulk pMF generated at the \mycomment{armchair-like }$x$ open edges as a result of the $\mathcal{T}$-invariant nature of pMF or more specifically, the abrupt cutoff of the spatial modulation giving rise to $\bm A$.

\section{Symmetric gauge}\label{App:gauge}
For pMF generated in a circuit, a gauge choice is physical although the low-energy behaviour around an exceptional degeneracy remains the same. Because of the reduced dimensionality, the Landau gauge proves more preferable for an easier circuit fabrication in size and modulation, a spectrum less complicated by the surface modes, and more accessible flat-band wavefunction towards detection. However, it is certainly worth inspecting the symmetric gauge case. We construct the wavefunctions in the continuum theory and compare with the circuit calculation. 
\begin{figure}[hbt]
\includegraphics[width=\columnwidth]{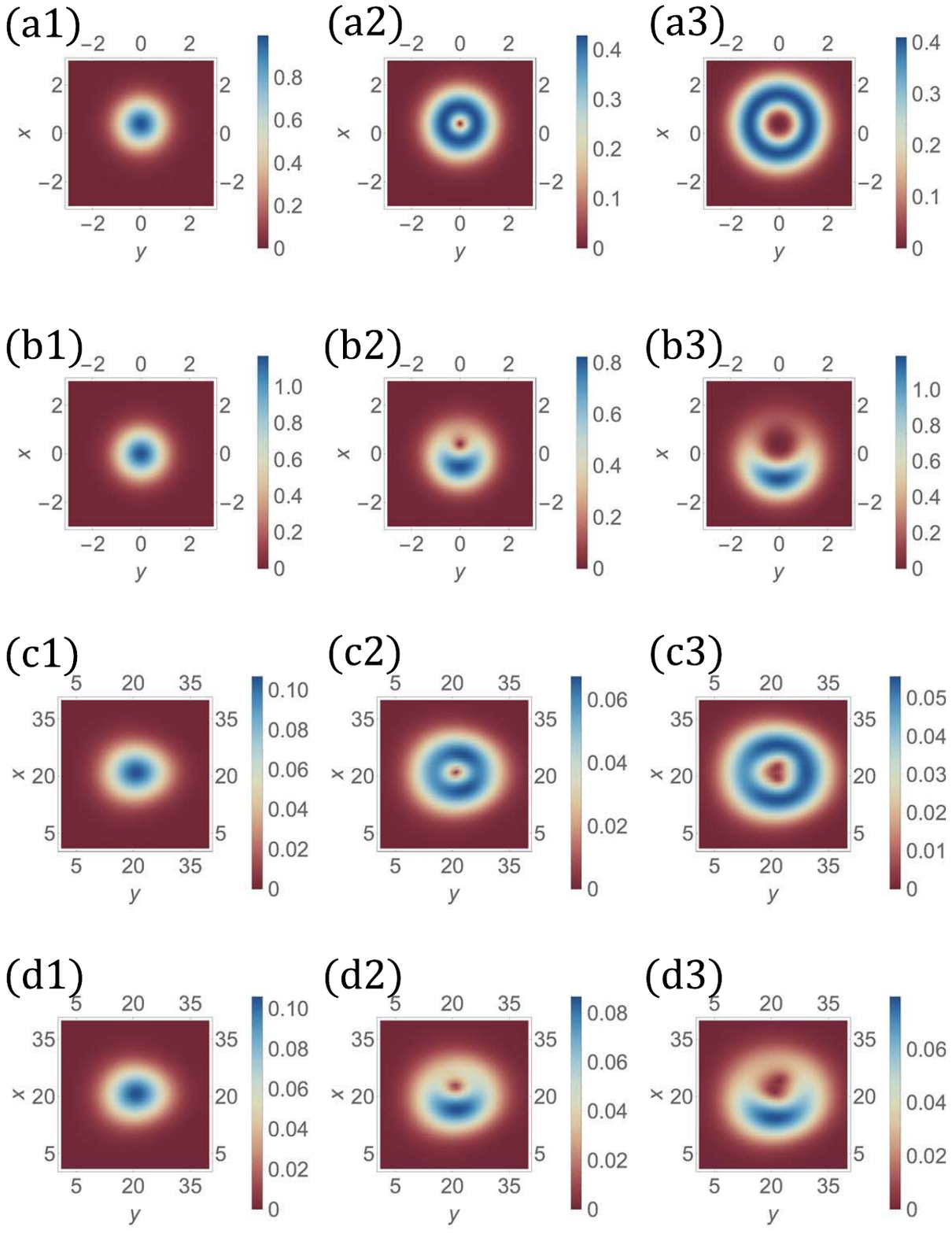}
\caption{Plots of the wavefunction amplitude under symmetric gauge pMF. a$n$) [b$n$)] is the $n$th analytic Hermitian (nH) ground state with conventional displacement $\delta=0.4+0\ii$ (with nH but same displacement $\delta$ due to nH $\gamma_x=\frac{b}{2} \Re\delta$ when $b=4$). c$n$) [d$n$)] is the $n$th lattice/circuit Hermitian (nH) sublattice-polarized chiral LL$_0$ state at $k_z=k_R$ on the dominant sublattice $A$ (with displacement due to nH $\gamma_x=0.06$ at $b=0.04$). }\label{Fig:gauge}
\end{figure}

With the symmetric gauge $\bm A=\frac{b}{2}(-y,x,0)$, it is 
convenient to set both $\gamma_x,\gamma_y$ nonzero. The Landau gauge formulation can be extended by choosing $\bm\epsilon=\frac{2}{b}\hat{z}\times\bm\gamma\mycomment{=2(-\gamma_y,\gamma_x,0)/b}$. To construct the nH ground state and coherent state, we work in the complex plane $\mathsf{z}=x+\ii y$ and most importantly, define a complex displacement $\delta=\frac{2}{b}(\gamma_x+\ii\gamma_y)$ due to non-Hermiticity. In contrast to the Hermitian case, we have independent variables  
\begin{equation}
w=\mathsf{z}-\delta,w'=\bar{\mathsf{z}} + \bar \delta    
\end{equation}
(and corresponding complex derivatives $\partial,\partial'$) that are not conjugate to each other. Note that $w$ exactly follows the physical meaning borne by $\bm\epsilon$. The merit of this construction is that two pairs of ladder operators of the Hermitian case remain formally intact, which read 
\begin{equation}
    B=-\ii\sqrt{2} ( l_b\partial' + \frac{w}{4l_b} )\,, B'=-\ii\sqrt{2} ( l_b\partial - \frac{w'}{4l_b} )
\end{equation}
 and 
\begin{equation}
\begin{split}
D&=(b/2)^{1/2}(X-\ii Y)=\sqrt{2} ( l_b\partial + \frac{w'}{4l_b} )\\ 
D'&=(b/2)^{1/2}(X+\ii Y)=\sqrt{2} ( -l_b\partial' + \frac{w}{4l_b} )
\end{split}
\end{equation}
where the center of cyclotron motion $(X,Y)$ is translated to the complex plane by $\bm\epsilon$. The unnormalized ground state 
\begin{equation}
    \ket{\psi_\mathrm{nHGS}}=(f(w)\ee^{-ww'/4l_b^2},0)^\mathrm{T}
\end{equation}
for any analytic function $f(w)$ and the coherent state 
\begin{equation}
\psi_\mathrm{nHCS}=\ee^{-(w-w_0)(w'-\bar w_0)/4l_b^2}     
\end{equation}
for any complex constant $w_0$.
Choosing the independent functions $f(w)$ as monomials $w^n$ by successively applying $D'$, we have $|\psi_\mathrm{nHGS}|^2=|w|^{2n}\ee^{-(|\mathsf{z}|^2+|\delta|^2)/2l_b^2}$. The envelop exponential function is circularly symmetric while the inner distribution becomes around the nH center $\delta$ as shown in Fig.~\ref{Fig:gauge}ab
. Although the probability 
$|\psi_\mathrm{nHCS}|^2$ remains a Gaussian packet centered at $w_0$, an extra phase structure is added by the nH effect. Also, $\psi_\mathrm{nHCS}$ reduces to a Hermitian ground state $\psi_\mathrm{CS}=\ee^{-\mathsf{z}(\bar{\mathsf{z}} + 2\bar\delta)/4l_b^2}$ when $w_0=-\delta$.
\begin{figure}[htb]
\includegraphics[width=0.8\columnwidth]{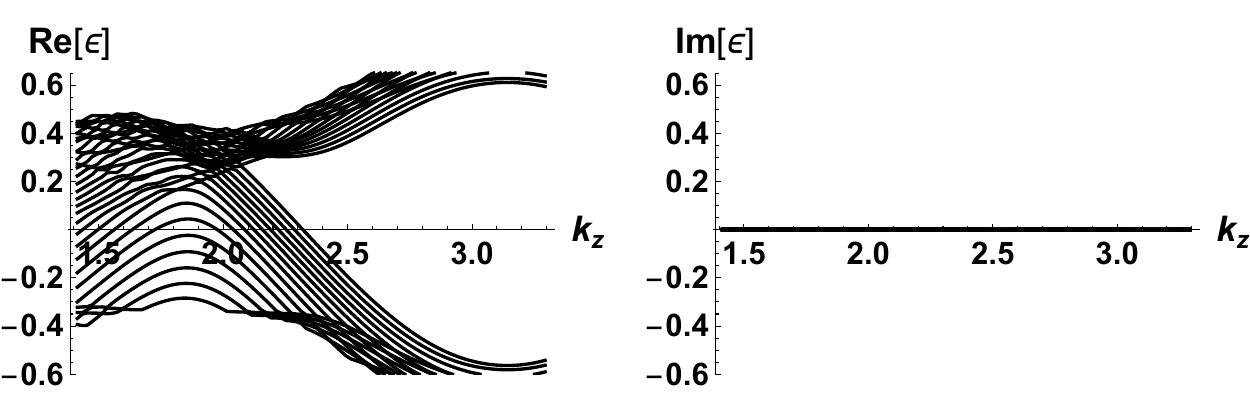}
\caption{Band structure of the nH chiral LL$_0$s around $k_z=k_R$ with vanishing imaginary part for the Weyl ER circuit under symmetric-gauge pMF $b\hat z$.}\label{Fig:symmkz}
\end{figure}

We thus choose the $n$th independent unnormalized nH ground state to be $\psi_\mathrm{nHGS}^{(n)}=(\varphi^{(n)},0)^\mathrm{T}$ where $\varphi^{(n)}=w^n\ee^{-ww'/4l_b^2}$. As a comparison in Fig.~\ref{Fig:gauge}ab, we also show the Hermitian case where $\varphi^{(n)}=w^n\ee^{-(z\bar z-2z\delta+\delta\bar\delta)/4l_b^2}$ with normal coordinate displacement of the same $\delta$. We make use of the model Eq.~\eqref{eq:Hsymm}, the Weyl ER circuit with four Weyl ER centers on the $z$-axis, to have a look at the symmetric gauge wavefunctions, which uses open boundary condition in $x,y$ and periodic connection in $z$. In its band structure shown in Fig.~\ref{Fig:symmkz}, the a few nH chiral LL$_0$s in the vicinity of $k_z=k_R$ correspond to the $\psi_\mathrm{nHGS}^{(n)}$'s. They are shown in Fig.~\ref{Fig:gauge}cd with the Hermitian case as a reference. One can clearly see the resemblance between Fig.~\ref{Fig:gauge}b and \ref{Fig:gauge}d.

\section{nH detection formalism}\label{App:detection}
In the biorthonormal representation, we have the spectral expansion of the impedence matrix $J^{-1}=\sum_{m}j_m^{-1}\ket{\tilde\psi_m^R}\bra{\tilde\psi_m^L}$ as the inverse of the admittance matrix $J\mycomment{=\ii \cal{Y}H}$. Here, $m$ signifies all quantum numbers in the 2D EP circuit, for instance, including the crystal momentum $k_x$ and band index $n$ in a circuit setting with periodic (open) boundary in $x$ ($y$) direction. Thus, the system bears a complete right (left) basis set $\ket{\tilde\psi_{k_xn}^{R(L)}}=\ket{\phi_{k_x}}\ket{\psi_{k_xn}^{R(L)}}$ where $\braket{x|\phi_{k_x}}=\ee^{\ii k_x x}/\sqrt{L}$ is the Bloch wavefunction along the periodic direction with circumference $L$. Diagonalizing the lattice Hamiltonian $H(k_x)$ ($H^\dag(k_x)$), we get the right (left) eigenstate $\ket{\bar\psi^R_n}$ ($\ket{\bar\psi^L_n}$) with eigenenergy $ E _n(k_x)$ ($ E _n^*(k_x)$). Biorthonormalization is given by $\ket{\psi_m^R}=\ket{\bar\psi_m^R}/\sqrt{\braket{\bar\psi_m^L|\bar\psi_m^R}},\ket{\psi_m^L}=\ket{\bar\psi_m^L}/\sqrt{\braket{\bar\psi_m^R|\bar\psi_m^L}}$. In the following, we use $(x,y,s)$ to denote a node's cell coordinate $(x,y)$ and sublattice location $s=A,B$. In the 3D ER circuit with Landau gauge, we simply append a good quantum number $k_z$ to $m$ (a periodic coordinate $z$) in the same manner as $k_x$ ($x$) in all the derivations. For brevity, we illustrate the 2D case only below.

The impedance between any two nodes $a=(x_1,y_1,s_1)$ and $b=(x_2,y_2,s_2)$ reads 
\begin{equation} 
\begin{split}
    Z_{ab}=&(J^{-1})_{aa}+(J^{-1})_{bb}-(J^{-1})_{ba}-(J^{-1})_{ab} \\
    =&\sum_{m} {j_m^{-1}(\tilde\psi_m^R(a)-\tilde\psi_m^R(b))(\tilde\psi_m^L(a)-\tilde\psi_m^L(b))^*}\\
    =&\sum_{k_x,n}  j^{-1}_{k_x,n} (\tilde\psi_{k_x,n}^R(x_1,y_1,s_1) -\tilde\psi_{k_x,n}^R(x_2,y_2,s_2))\\
    &\times(\tilde\psi_{k_x,n}^L(x_1,y_1,s_1)-\tilde\psi_{k_x,n,s}^L(x_2,y_2,s_2))^* \\
    =&\sum_{k_x,n}  j^{-1}_{k_x,n} (\psi_{k_x,n}^R(y_1,s_1)-\psi_{k_x,n}^R(y_2,s_2) \ee^{\ii k_x(x_2-x_1)})\\
    &\times(\psi_{k_x,n}^L(y_1,s_1)-\psi_{k_x,n}^L(y_2,s_2) \ee^{\ii k_x(x_2-x_1)} )^* 
\end{split}
\end{equation}
The basis set $\ket{\tilde\psi_{k_xn}^R}$ can be used to expand the current injecting state $\ket{\Psi_i}=\sum_{k_xn}c_{k_xn}\ket{\tilde\psi_{k_xn}^R}$ 
with $c_{k_xn}= \bra{\phi_{k_x}} \braket{\psi_{k_xn}^L | \Psi_i} = \sum_{x,y,s}{ \frac{1}{\sqrt{L}} \ee^{-\ii k_xx} \psi^{L*}_{k_xn}(y,s) \Psi_i(x,y,s) }$. Then the voltage response state is
\begin{equation}
\begin{split}
    \ket{\Psi_v} &= J^{-1} \ket{\Psi_i}\\
    &= \sum_{k_xn} {c_{k_xn} \ket{\phi_{k_x}} \sum_{k_x'n'} { j_{k_x'n'}^{-1} \ket{\tilde\psi_{k_x'n'}^R} \braket{\tilde\psi_{k_x'n'}^L|\tilde\psi_{k_xn}^R} }  }\\
    &= \sum_{k_xn} {c_{k_xn} \ket{\phi_{k_x}} \sum_{n'} { j_{k_xn'}^{-1} \ket{\psi_{k_xn'}^R} \braket{\psi_{k_xn'}^L|\psi_{k_xn}^R} }  }\\
    &= \sum_{k_xn} {\frac{c_{k_xn}}{j_{k_xn}} \ket{\phi_{k_x}} \ket{\psi_{k_xn}^R} }  
\end{split}
\end{equation}
and the voltage at a particular node $(x,y,s)$ follows
\begin{equation}
\begin{split}
\Psi_v(x,y,s) = \sum_{k_xn} {\frac{c_{k_xn}}{j_{k_xn}} \frac{1}{\sqrt{L}}\ee^{\ii k_xx} \psi_{k_xn}^R(y,s) } .
\end{split}
\end{equation}
If we consider injecting current only to a specific node at the midpoint of the edge of the open direction, say, $(x_0,y_0,s_0)$, we have $\Psi_i(x,y,s)=\delta_{xx_0}\delta_{yy_0}\delta_{ss_0}$ and hence $c_{k_xn}=\frac{1}{\sqrt{L}}\ee^{-\ii k_xx_0}\psi^{L*}_{k_xn}(y_0,s_0)$. Therefore,
\begin{equation}
\begin{split}
\Psi_v(x,y,s) = \sum_{k_xn} {\frac{\psi^{L*}_{k_xn}(y_0,s_0)}{Lj_{k_xn}} \ee^{\ii k_x(x-x_0)} \psi_{k_xn}^R(y,s) }.
\end{split}
\end{equation}
On the other hand, to observe a specific nH wavefunction, one has to adopt a different injecting current, which is an eigenstate of a particular ac driving pattern of wave number $k_x'$, i.e., $\Psi_i'=\frac{1}{\sqrt{L}} \ee^{\ii k_x'x} \delta_{yy_0}\delta_{ss_0}$. Therefore, we have the expansion coefficient modified to $c_{k_xn}'=\delta_{k_xk_x'}\psi^{L*}_{k_xn}(y_0,s_0)$ and thus \begin{equation}
\begin{split}
\Psi_v'(x,y,s) = \sum_{n} {\frac{\psi^{L*}_{k_x'n}(y_0,s_0)}{\sqrt{L}j_{k_x'n}} \ee^{\ii k_x'x} \psi_{k_x'n}^R(y,s) }.
\end{split}
\end{equation}





\section{Circuit for nH Hofstadter butterfly}
\begin{figure}[htb]
\includegraphics[width=0.6\columnwidth]{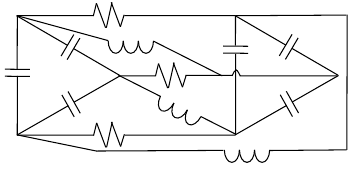}
\caption{Building block of a nH circuit simulating $\frac{1}{3}$ magnetic flux quantum per unit cell.}\label{Fig:Hofstadter}
\end{figure}
The nH pMF LL realized in this study corresponds, in fact, to the weak-field limit where the magnetic length $l_b$ is large compared to the lattice spacing. The complementary strong-field situation where $l_b$ is comparable to the lattice spacing leads to the Hofstadter butterfly. A natural question is what new features a nH Hofstadter butterfly would have. Here, we briefly point out a circuit realization towards exploring this for future study. 

The first step is to make lattice magnetic flux large enough. Instead of our pMF method, one can make a conventional circuit lattice model, however, with each node replaced by $q$ capacitors (inductors) connected as a ring or connected to a single node, either of which also provides $q$ outward subnodes \cite{Albert2015,Zhao2018}. In Fig.~\ref{Fig:Hofstadter} we adopt the ring setting with three internal capcitors. 
Each composite node acquires $q$ internal eigenmodes labeled by $k$ and most importantly, connecting corresponding subnodes by inductors (capacitors) with a cyclic shift $p$ between two such composite nodes will generate hoppings between these two composite main nodes with a gauge factor $\ee^{\ii2\pi k\frac{p}{q}}$ where $k,p=0,\cdots,q-1$. Then one can introduce non-Hermiticity by additionally connecting corresponding subnodes in two composite nodes by resistors in the same manner. When the shift $p=0$ for all resistive connections of identical resistance $R$, the non-Hermiticity is equivalent to the one in the main text. Otherwise, it would be a more complex case attaching gauge flux to the nH hoppings. In Fig.~\ref{Fig:Hofstadter}, we show the case of $q=3$ internal modes, $p=1$ for inductor connection and $p=0$ for nH resistor connection.

\end{document}